\documentclass[12pt,preprint]{aastex}

\shorttitle{X-ray Structure of Distant Galaxy Clusters}
\shortauthors{Ota \& Mitsuda}

\newcommand{\Fig}{Figure~}
\newcommand{\rc}{r_{\rm c}}
\newcommand{\kT}{{\rm k}T}
\newcommand{\LX}{L_{\rm X}}

\begin{document}

\title{X-ray Study of Seventy-nine Distant Clusters of Galaxies:
Discovery of Two Classes of Cluster Size}

\author{Naomi Ota\altaffilmark{1,2,3} and
Kazuhisa Mitsuda\altaffilmark{3}}

\altaffiltext{1}{Department of Physics, Tokyo Metropolitan University, 
1-1 Minami-osawa, Hachiouji, Tokyo 192-0397, Japan}
\altaffiltext{2}{naomi@phys.metro-u.ac.jp}
\altaffiltext{3}{Institute of Space and Astronautical Science, 
Sagamihara, Kanagawa 229-8510, Japan}

\begin{abstract} 
We have performed a uniform analysis of 79 clusters of galaxies with
the ROSAT HRI and ASCA to study the X-ray structure and evolution of
clusters in the redshift range $0.1<z<1$. We determined the average
X-ray temperatures and the bolometric luminosities with ASCA and the
spatial distributions of the X-ray brightness with the ROSAT HRI by
utilizing the isothermal $\beta$-model. We do not find any significant
redshift dependence in the X-ray parameters including the temperature,
$\beta$-model parameters, and the central electron density. Among the
parameters, the core radius shows the largest cluster-to-cluster
dispersions. We discovered that the histogram of the core radius
shows two peaks at 60 and 220 kpc. If we divide the cluster samples
into two subgroups corresponding to the two peaks in the core radius
distribution, they show differences in the X-ray and optical
morphologies and in the X-ray luminosity-temperature relation.  From
these observational results, we suggest that the clusters are divided
into at least two subgroups according to the core radius.
\end{abstract}


\section{Introduction}
Clusters of galaxies are the largest collapsed structures in the
universe, and thus are a useful cosmological probe. At X-ray energies,
thermal emission from highly ionized plasma that is confined in the
cluster is an excellent tracer of the underlying gravitational
potential. A large number of cluster samples observed with the ASCA
and ROSAT observatories enable us to investigate both the spatial and
spectral structures systematically. \cite{MOHR_ETAL.1999} performed a
uniform analysis on the ROSAT PSPC data of forty-five clusters with $z
^{<}_{\sim} 0.1$ under the $\beta$-model \citep{Cavaliere_Fusco_1976}
and gathered the spectral data from the literature. Based on their
results,
\cite{FUJITA_ETAL.1999a} found that the nearby clusters show a planar
distribution in the 3-dimensional parameter space, and named it the
X-ray fundamental plane. \cite{FUJITA_ETAL.1999b} proposed that the
major axis of the plane may represent the cluster formation epoch.
According to their interpretations, the distributions of more distant
clusters on the plane should be displaced in comparison to the nearby
samples.

In order to constrain structure and evolution of clusters, it is
important to extend the cluster study to a wider redshift
range. However, at $z>0.1$, only a limited number of clusters have
been uniformly analyzed
(e.g. \cite{ALLEN.1998,ETTORI_ETAL.1999,HASHIMOTODANI.1999}). In this
paper and \cite{OTA.2001}, we analyzed the ROSAT HRI and the ASCA
SIS/GIS data of 79 clusters in a uniform manner for the first
time. The analysis includes the largest sample of pointed observations
of distant clusters. We report on the results of the redshift
dependence of the X-ray properties and the double structure discovered
in the spatial distributions of the intracluster gas. We use $H_0= 50$
km/s/Mpc, $\Omega_{\rm M} = 0.3$, and $\Omega_{\Lambda} = 0.7$
throughout the paper.

\section{Systematic analysis}

\subsection{The sample} 
We have selected distant clusters with $0.1 < z < 1$ for which both
ASCA and the ROSAT HRI pointed observations were made. Excluding three
clusters that were observed away from the center of the field of view
of the ROSAT HRI, the sample comprises 79 clusters. Clusters with
$z<0.4$ make up about 90\% of the sample.  Since the samples are
collected from proposal observations, and also since the sensitivities
for high-redshift clusters are limited, we have to be careful about
selection bias.  Our sample covers the temperature range almost
equivalent to that of the X-ray flux limited 55 cluster sample
constructed by \cite{EDGE_ETAL.1990}, but has a higher average
temperature of 6.8 keV. The K-S test shows the probability that the
two samples are from the same temperature distribution is 0.06 (the
K-S parameter, $D = 0.24$).  Observation bias will be discussed in a
later section in more detail.

\subsection{Spatial analysis}

We have made use of the ROSAT Data Archive of the Max-Plank-Institut
f\"{u}r extraterrestrische Physik to study the spatial
distributions. We used the EXSAS analysis package
\citep{ZIMMERMANN_ETAL.1992} to produce the HRI images from the event
lists and rebinned the images into $5''$ bins. We first exclude image
regions of point sources that were detected by the maximum likelihood
method. Then we define the cluster center as the center of gravity of
the photon distribution in a given aperture radius, and derive
azimuthally averaged radial profiles. We fit the radial profiles with
the isothermal $\beta$-model to determine the best-fit model
parameters that minimize $\chi^2$. The single $\beta$-model fitting
function is written as $ S(r) = S_0 ( 1 + (r/\rc)^2 )^{-3\beta+1/2} +C
$, where $S_0$, $\rc$, and $\beta$ are the central surface brightness,
the core radius and the outer slope, respectively, and $C$ is a
constant background.

In order to estimate the possible systematic errors, we studied the
dependence of the best-fit parameter values on analysis parameters;
the aperture radius for the cluster centroid determination, and the
inner and outer radii of the image region used in the model fittings.

The centroid positions of 34 clusters showed significant (more than
$3\sigma$) variations as a function of the aperture radius of the
centroid determination compared to the statistical errors estimated by
Monte-Carlo simulations. We find the variation is due to irregularity
of the photon distribution. Thus we define the 34 clusters as
``irregular'' clusters, and the others as ``regular'' clusters.
However, even for the irregular clusters, the deviations of the
best-fit $\beta$-model parameters due to the selection of the aperture
radius were smaller than the statistical errors, as long as the
aperture radius is large enough that there is no significant cluster
emission outside it but it is smaller than $12'$, within which radial
dependences of telescope vignetting and background are small
\citep{Snowden_1998}.

If the outer radius used for the fitting is too small, we overestimate
or underestimate the background level and introduce systematic errors
in the fitting parameters.  However, we find the best-fit values are
constant within the statistical errors as long as the outer radius is
large enough, but $^{<}_{\sim} 12'$.

The fit significantly improved for nine ``regular'' clusters when we
excluded a few central bins from the fitting. We restricted this
analysis to the regular clusters to avoid apparent double structure
that could arise from the irregularities.  The best-fit core radius
increased for seven of them when we excluded the central bins. This
indicates that those clusters contain two different length
scales. Thus we define the double $\beta$-model as a superposition of
two single $\beta$-models with different core radii. For all the nine
clusters, the radial profiles are better fitted with the double
$\beta$-model than the single $\beta$-model; the improvement is
significant by the F-test at the 95\% level. We refer to those nine
clusters as the double-$\beta$ clusters hereafter. Of these, we
classify seven clusters as inner-core dominant double-$\beta$, because
the central surface brightness of the inner-core component is higher
by one to two orders of magnitude than that of the outer-core
component. We call the remaining two clusters outer-core dominant
double-$\beta$, because the outer-core component dominates the total
luminosity.

\subsection{Spectral analysis}

We integrated the cluster spectra from the circular regions of radius
$3'$ for the SIS and $6'$ for the GIS. We subtracted the background
spectra that were obtained through blank sky observations. Assuming a
thin-thermal emission model \citep{RAYMOND_ETAL.1977}, we fitted the
SIS and GIS spectra simultaneously, where the absorption column
density was allowed to vary in order to take into account the
uncertainty in the low energy efficiency of the SIS detector
\citep{YAQOOB.1999}. We determined the emission-weighted average
temperature, $\kT$ and also estimated the bolometric luminosity,
$L_{\rm X, bol}$.  We evaluated systematic errors caused by the
selection of integration area and the background subtraction in the
analysis, and found they are smaller than the statistical errors.  We
checked the consistency of the luminosities from ASCA and ROSAT. We
find the luminosity from the HRI surface brightness profile is
systematically higher by about 15\%. We consider that this is within
calibration errors of the effective areas of the X-ray telescopes 
(\cite{FUKAZAWA_ETAL.1997}; \cite{PRESTWITCH_ETAL.1998})

\section{Results}

\subsection{Redshift dependence}

\begin{figure} 
\plotone{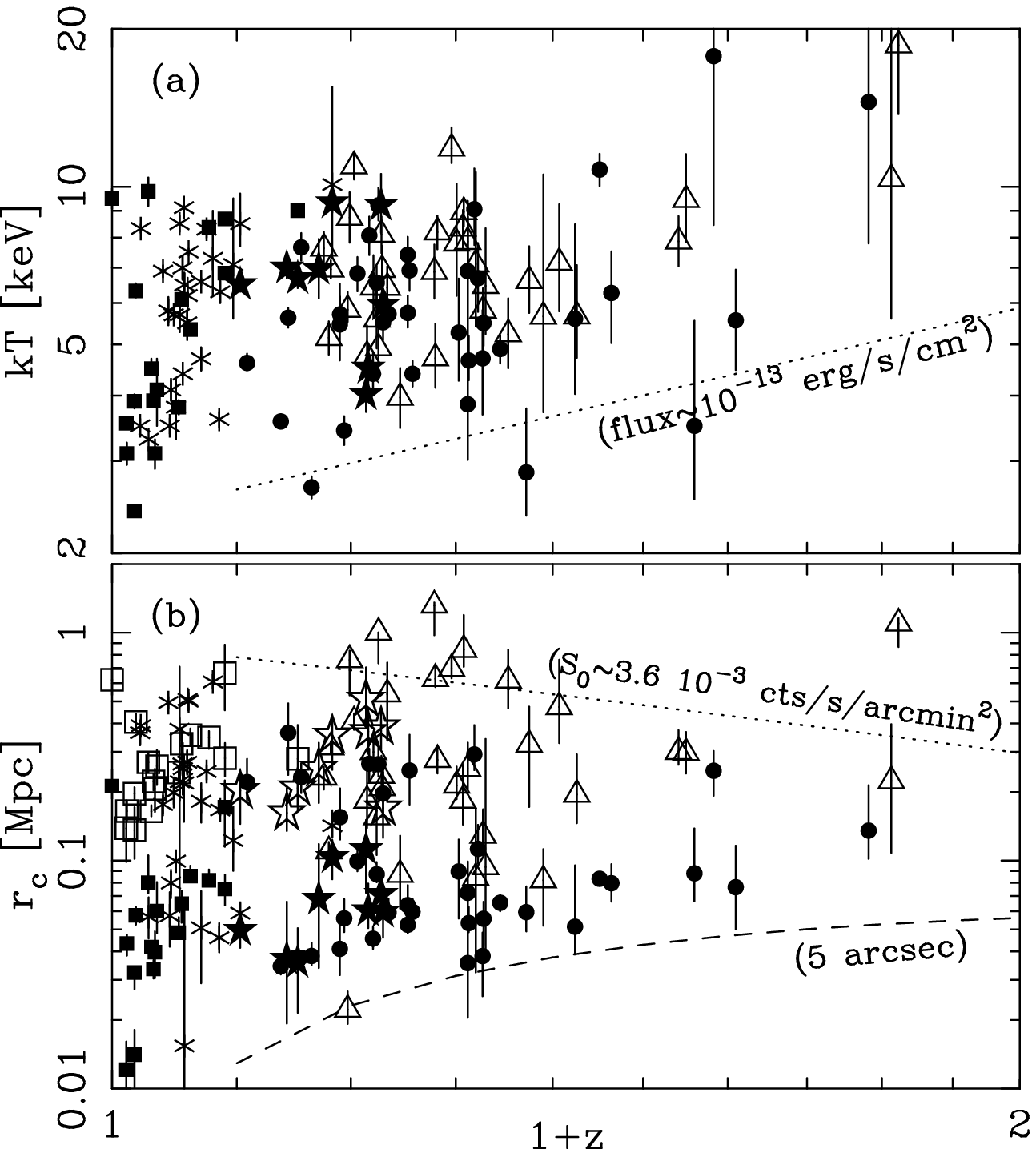} 
\caption{Redshift dependence of the average temperature (a) and the
core radius (b). Subgroups of clusters are denoted with different
symbols, the single $\beta$ regular and irregular clusters (circles
and triangles), the double $\beta$ clusters (stars) where the core
radii of two components are plotted with open and filled stars in
panel (b), the single and double $\beta$ clusters of nearby samples by
\cite{MOHR_ETAL.1999} (asterisks and open/filled squares). 
The errors are at 90\% confidence.  The upper curve in panel (b)
corresponds to a constant central surface brightness of $S_0
=3.6\times10^{-3} \,{\rm counts/s/arcmin^2}$, which is the typical
sensitivity of the present observations. In order to draw the curve,
we assumed $\beta=0.7$ and $\kT = 5$ keV.
\label{fig1}} 
\end{figure}

In \Fig\ref{fig1}, we plotted the redshift dependence of the average
temperature and core radius obtained from the present analysis
together with the data for clusters with $z^{<}_{\sim}0.1$ from
\cite{MOHR_ETAL.1999}. With three overlaps there are 121 clusters. We
do not find many clusters with $\kT < 5$ keV at $z > 0.4$. As
indicated in \Fig\ref{fig1} (a), this is consistent with the ASCA
sensitivity, if we assume the luminosity-temperature relation of the
clusters (See subsection \ref{subsec:Morpho_LT}). We also do not find
many clusters with a core radius larger than $\sim 300$ kpc or smaller
than $\sim 40$ kpc at $z > 0.4$. This is again consistent with the
selection effect, because the detection of clusters with very small
and large core sizes are limited respectively by the spatial
resolution ($5''$) and by the sensitivity for low surface brightness
emission (see sensitivity curves in \Fig\ref{fig1} (b)). Therefore,
due to the sensitivity limit, the redshift evolution of the
temperature and the core radius is not clear, although there is large
cluster-to-cluster scatter. We also found that $\beta$ and the central
electron density have no significant redshift dependence either. The
average and the standard deviation of $\beta$ for the 79 samples are
0.65 and 0.29, respectively.

It is remarkable that the core radius shows the largest
cluster-cluster dispersions among the X-ray parameters, spanning over
an order of magnitude. We also notice in \Fig\ref{fig1} (b) that the
number of clusters with $\rc \sim 100$ kpc is smaller compared to
clusters with $\rc \sim$ 50 and 200 kpc.

The fraction of the irregular clusters in our sample is 43\%. This is
smaller than the fractions of irregular clusters of nearby clusters;
71\% for 46 Einstein IPC samples \citep{MOHR_ETAL.1995}, 49\% of 39
ROSAT PSPC samples (result with a $1\,h_{80}^{-1}$Mpc aperture in
\cite{BUOTE_ETAL.1996}). All the substructure tests 
measure departures from regularity against the level expected from
Poisson noise. The typical number of photons of our cluster sample is
by a factor of $\sim 2$ smaller than that of the nearby sample of 
\cite{MOHR_ETAL.1995}. Thus this may partly account for the
difference. Difference in analysis techniques to quantify
irregularities and selection bias may also partly explain the
difference.

\subsection{Core radius distribution}

\begin{figure}
\plotone{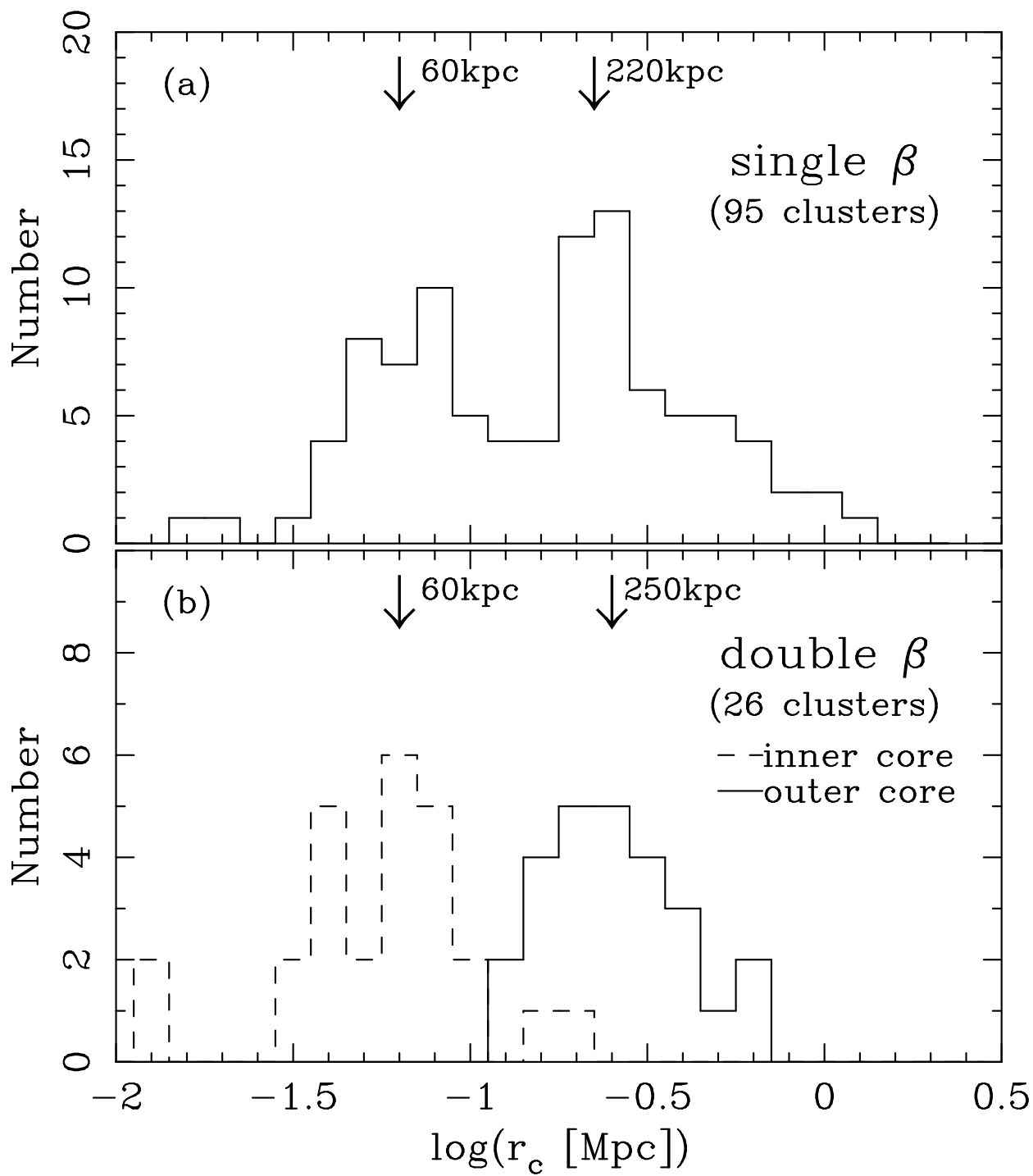}
\caption{Histograms of the core radius for the single-$\beta$ clusters
(a) and the double-$\beta$ clusters (b). Our distant ($z>0.1$) sample
and the nearby sample of \cite{MOHR_ETAL.1999} are added together. For
the double-$\beta$ clusters, the inner and the outer components are
plotted separately with the solid and dashed lines. \label{fig2}}
\end{figure}

In \Fig\ref{fig2} we show histograms of the core radius for the
single-$\beta$ and the double-$\beta$ clusters, separately.  In the
histograms we included the nearby clusters of 
\cite{MOHR_ETAL.1999}. Eighteen of their samples are better fitted
with the double $\beta$-model.  For the single-$\beta$ clusters, there
are clearly two peaks at 60 kpc and 220 kpc.  They are separated by
about a factor of 4 and the depletion around 100 kpc is significant at
the $4\sigma$ level if the population is comparable to those in the
two peaks. About 70\% of clusters (64 out of 95) are within two
core-radius ranges corresponding to the peaks: $ 40 < \rc < 80 $ kpc
and $160 < \rc < 320$ kpc.  Although the detection of large and small
core clusters are seriously affected by the selection effect at
$z>0.4$, there is no reason that makes the detection of the
intermediate core sizes of $\sim100$ kpc difficult and it is very
unlikely that the proposers of those observations avoided clusters of
the intermediate core size. Therefore, as the observed double-peaked
distribution cannot be explained by selection effects, it reflects the
real nature of the clusters. For the double-$\beta$ clusters, the
inner core and the outer core are distributed around 60 kpc and 250
kpc, respectively. The coincidence with the two peak values of the
single-$\beta$ clusters is also striking.

\subsection{Optical morphology and $\LX-T$ relation}
\label{subsec:Morpho_LT}
In \Fig\ref{fig1}, we find a systematic difference in $\rc$ between
the regular and the irregular clusters; clusters with a small (large)
core tend to have a regular (irregular) X-ray morphology. Furthermore,
among 37 clusters for which the optical Bautz-Morgan type is known,
all 6 clusters classified as BM I or I-II (i.e. a cluster with a
cD galaxy) are found to be either a single-$\beta$ cluster with a
small core or a double-$\beta$ cluster.

Concentrated gas distribution in the small core system is indicated
from the plot of the $\LX-T$ relation (\Fig\ref{fig3}). The clusters
with $\rc < 135$ kpc and $>135$ kpc are denoted by different
symbols.  The core radius of the dominant core was used for the
classification of the double-$\beta$ clusters. We obtained the
observed $\LX-T$ relations of the form $\log{\kT} = a (\log{L_{\rm X,
bol}} - 45.4) + b$ with $(a,b)=(0.37\pm0.12, 0.68\pm0.04)$ for $\rc <
135$ kpc and $(a,b)=(0.324\pm0.097, 0.81\pm0.03)$ for $\rc >135$ kpc,
where the errors (90\%) were estimated from the dispersion of the data
points around the model function rather than the photon
statistics. Thus the slopes are consistent for the two subgroups,
however, the offsets differ significantly. The difference is
consistent with a higher central electron density for the small core
group: we find from the $\beta$-model analysis that $n_{e0}$ is
approximately related to $\rc$ through $n_{e0} \propto \rc^{-\alpha}$
with $\alpha \sim 1.5$. \cite{FABIAN_ETAL.1994} mentioned that the
offset of clusters from the mean $\LX-T$ relation can be explained by
the strength of cooling flow. We discuss the connection between the
double peak structure of the core radius distribution and the $\LX-T$
relation in more detail in a separate paper \citep{Mitsuda_Ota_2002}.

\section{Discussion}
We found that among the X-ray parameters, the core radius has the
largest scatter and that its histogram shows a double-peaked
distribution.  The optical and X-ray morphologies
and the $\LX-T$ relation show correlations with
the core radius. These results indicate that the clusters
are divided into two subgroups according to the core radius.

In some nearby clusters, a central concentration in the iron
distribution has been detected (e.g. \cite{EZAWA_ETAL.1997}). This may
enhance the central surface brightness and might produce a
double-$\beta$ profile. We generated simulated single-$\beta$ cluster
images assuming the abundance gradient profile observed for AWM7 in
which the abundance increases from 0.2 to 0.5 solar toward the center,
and the average temperature from 0.5 to 3 keV. When the temperature is
3 keV, the central excess emission attributed to the gradient is only
about 10\% and the radial profile is well fitted with the single
$\beta$-model.  When $\kT$ is lower than 1 keV, the radial profile
approaches that of the outer core dominant double-$\beta$ cluster.
Thus the outer core dominant double-$\beta$ clusters can be explained
by simultaneous abundance and temperature gradients, because although
the observed average temperatures are all higher than 2 keV, the
temperature of the inner core region can be lower for outer core
dominant double-$\beta$ clusters.  However, the inner-core dominant
double-$\beta$ clusters cannot be explained with this model.

Because of the similarity of the core radius distributions between the
single-$\beta$ and the double-$\beta$ clusters, we tested the
assumption that the clusters classified as the single-$\beta$ regular
clusters also contain a double-$\beta$ profile within the photon
statistics, assuming the ratio of the two core radii to be 1 : 4, and
the ratio of central surface brightnesses 1 : 0.01 to 1 : 1. As a
result, we found that the double $\beta$-model cannot be rejected at
the 90 \% level for about 2/3 of the single-$\beta$ regular clusters.
Thus it is possible that a large fraction of the clusters contain two
different length scales in X-ray brightness distribution and can be
approximated with the double $\beta$-model. The fraction of
double-$\beta$ clusters in nearby regular clusters
\citep{MOHR_ETAL.1995,MOHR_ETAL.1999} is $\sim60$\% (4 out of 7
regular clusters). Thus the fractions are consistent for the
distant and the nearby samples.

\begin{figure} 
\plotone{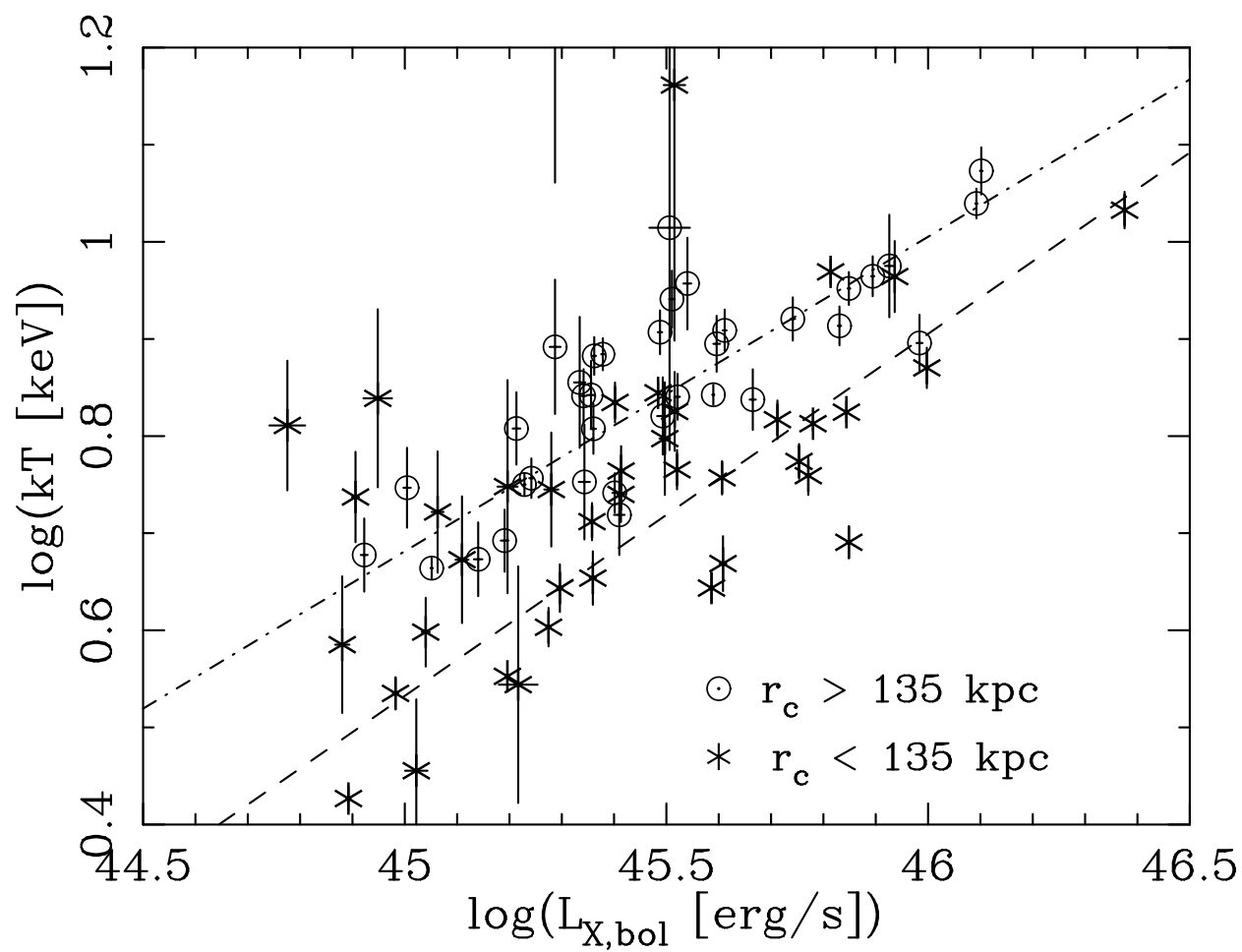} 
\caption{Luminosity-temperature
relations derived from the ASCA observations of the distant
clusters. The asterisks and the circles denote the clusters with
$\rc<135$ and $\rc>135$ kpc, respectively. The vertical and horizontal
error bars are the $1\sigma$ errors. The best-fit power-law functions
are shown with the dashed and dot-dashed lines for the two subgroups.
\label{fig3}} 
\end{figure}

The radiative cooling timescale at the cluster center is shorter than
the age of the universe for most of the small core clusters.  This
supports the presence of the cooling flow phenomenon (e.g.
\cite{Fabian_1994}). The cooling flow model predicts the emergence of a
cool component at the cluster center.  If the small core radius of the
small-core subgroup is due to such a cool component, these clusters
should contain another emission component of a larger core radius
which reflects the gravitational potential. From the upper limit of
the surface brightness of the outer component from the analysis in the
previous paragraph, we find the photon flux from the outer component
is smaller than 25\% of the total emission for most of clusters of the
subgroup. Thus the average cluster temperature should reflect the
average temperature of the central enhancement.  However we do not
observe a strong correlation between $\rc$ and $\kT$. In particular,
the temperature ranges of clusters with $\rc <135$ and $\rc > 135$ kpc
are comparable to each other (see \Fig\ref{fig3}). The recent
XMM-Newton observations of nearby cooling flow clusters revealed that
the temperature gradient is not very large and is approximately
$r^{-0.2}$ (\cite{BOHRINGER_ETAL.2001,TAMURA_ETAL.2001}). These
indicate that cooling is not as efficient as the standard cooling flow
predictions. However a small temperature gradient of the gas could cause
differences in the gas density distribution and the underlying
potential distribution. We need further investigation to clarify
through which physical processes such discrete cluster structures are
formed.

\acknowledgments 
We are grateful to M. Hattori and S. Sasaki for useful comments.  We
thank Philip Edwards for his careful review of the manuscript. N.O.
is supported by a Research Fellowship for Young Scientists from the
JSPS.

\clearpage

\end{document}